\documentclass[twoside,twocolumn,9pt]{article}
\usepackage{extsizes}
\usepackage[super,sort&compress,comma]{natbib} 
\usepackage[version=3]{mhchem}
\usepackage[left=1.5cm, right=1.5cm, top=1.785cm, bottom=2.0cm]{geometry}
\usepackage{balance}
\usepackage{comment}
\usepackage{mathptmx}
\usepackage{sectsty}
\usepackage{graphicx} 
\usepackage{lastpage}
\usepackage[format=plain,justification=justified,singlelinecheck=false,font={stretch=1.125,small,sf},labelfont=bf,labelsep=space]{caption}
\usepackage{float}
\usepackage{fancyhdr}
\usepackage{fnpos}
\usepackage[english]{babel}
\addto{\captionsenglish}{%
  
}
\usepackage{array}
\usepackage{droidsans}
\usepackage{charter}
\usepackage[T1]{fontenc}
\usepackage[usenames,dvipsnames]{xcolor}
\usepackage{setspace}
\usepackage[compact]{titlesec}
\usepackage{hyperref}
\usepackage{amsmath}
\usepackage{soul}

\usepackage{epstopdf}
\usepackage{multirow}
\usepackage{multicol}

\definecolor{cream}{RGB}{222,217,201}

\begin{document}

\pagestyle{fancy}
\thispagestyle{plain}
\fancypagestyle{plain}{
\renewcommand{\headrulewidth}{0pt}
}

\makeFNbottom
\makeatletter
\renewcommand\LARGE{\@setfontsize\LARGE{15pt}{17}}
\renewcommand\Large{\@setfontsize\Large{12pt}{14}}
\renewcommand\large{\@setfontsize\large{10pt}{12}}
\renewcommand\footnotesize{\@setfontsize\footnotesize{7pt}{10}}
\makeatother

\renewcommand{\thefootnote}{\fnsymbol{footnote}}
\renewcommand\footnoterule{\vspace*{1pt}%
\color{cream}\hrule width 3.5in height 0.4pt \color{black}\vspace*{5pt}} 
\setcounter{secnumdepth}{5}

\makeatletter 
\renewcommand\@biblabel[1]{#1}            
\renewcommand\@makefntext[1]%
{\noindent\makebox[0pt][r]{\@thefnmark\,}#1}
\makeatother 
\renewcommand{\figurename}{\small{Fig.}~}
\sectionfont{\sffamily\Large}
\subsectionfont{\normalsize}
\subsubsectionfont{\bf}
\setstretch{1.125} 
\setlength{\skip\footins}{0.8cm}
\setlength{\footnotesep}{0.25cm}
\setlength{\jot}{10pt}
\titlespacing*{\section}{0pt}{4pt}{4pt}
\titlespacing*{\subsection}{0pt}{15pt}{1pt}

\fancyfoot{}
\fancyfoot[LO,RE]{\vspace{-7.1pt}\includegraphics[height=9pt]{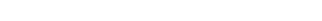}}
\fancyfoot[CO]{\vspace{-7.1pt}\hspace{13.2cm}\includegraphics{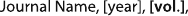}}
\fancyfoot[CE]{\vspace{-7.2pt}\hspace{-14.2cm}\includegraphics{head_foot/RF}}
\fancyfoot[RO]{\footnotesize{\sffamily{1--\pageref{LastPage} ~\textbar  \hspace{2pt}\thepage}}}
\fancyfoot[LE]{\footnotesize{\sffamily{\thepage~\textbar\hspace{3.45cm} 1--\pageref{LastPage}}}}
\fancyhead{}
\renewcommand{\headrulewidth}{0pt} 
\renewcommand{\footrulewidth}{0pt}
\setlength{\arrayrulewidth}{1pt}
\setlength{\columnsep}{6.5mm}
\setlength\bibsep{1pt}

\makeatletter 
\newlength{\figrulesep} 
\setlength{\figrulesep}{0.5\textfloatsep} 

\newcommand{\topfigrule}{\vspace*{-1pt}%
\noindent{\color{cream}\rule[-\figrulesep]{\columnwidth}{1.5pt}} }

\newcommand{\botfigrule}{\vspace*{-2pt}%
\noindent{\color{cream}\rule[\figrulesep]{\columnwidth}{1.5pt}} }

\newcommand{\dblfigrule}{\vspace*{-1pt}%
\noindent{\color{cream}\rule[-\figrulesep]{\textwidth}{1.5pt}} }

\makeatother


\twocolumn[
  \begin{@twocolumnfalse}
{\includegraphics[height=30pt]{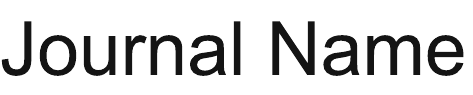}\hfill\raisebox{0pt}[0pt][0pt]{\includegraphics[height=55pt]{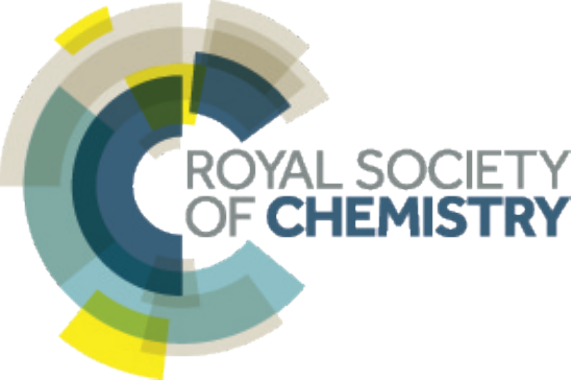}}\\[1ex]
\includegraphics[width=18.5cm]{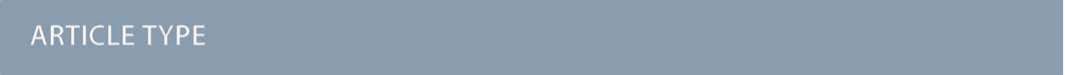}}\par
\vspace{1em}
\sffamily
\begin{tabular}{m{4.5cm} p{13.5cm} }

\includegraphics{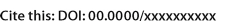} & \noindent\LARGE{\textbf{ First-Principles Calculations on Monolayer \ce{WX2} (X = S, Se) as an Effective Drug Delivery Carrier for Anti-Tuberculosis Drugs$^\dag$}} \\
\vspace{0.3cm} & \vspace{0.3cm} \\

 & \noindent\large{Khaled Mahmud,\textit{$^{a}$}$^{\ast}$ Taki Yashir, \textit{$^{a}$}$^{\ast}$ and Ahmed Zubair\textit{$^{a\ddag}$} } \\

\includegraphics{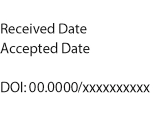} & \noindent\normalsize{Tuberculosis (TB) remains a major global health concern, necessitating the exploration of novel drug delivery systems to combat the challenges posed by conventional approaches. We investigated the potential of monolayer transition metal dichalcogenides (TMDs) as an innovative platform for efficient and targeted delivery of antituberculosis drugs. Specifically, the electronic and optical properties of prominent TB drugs, isoniazid (INH) and pyrazinamide (PZA), adsorbed on tungsten diselenide (WSe$_{2}$) and tungsten disulfide (WS$_{2}$) monolayers were studied using first-principles calculations based on density functional theory (DFT). The investigation revealed that the band gaps of WSe$_{2}$ and WS$_{2}$ monolayers remain unaltered upon adsorption of PZA or INH, with negative adsorption energy indicating stable physisorption. We explored different vertical and horizontal configurations, and the horizontal ones were more stable. When INH and PZA drugs were horizontally adsorbed together on WSe$_{2}$, the most stable configuration was found with an adsorption energy of -2.35 eV. Moreover, the adsorbed drugs could be readily released by light within the visible or near-infrared (NIR) wavelength range. This opened up possibilities for their potential application in photothermal therapy, harnessing the unique properties of these 2D materials. The comprehensive analysis of the band structures and density of states provides valuable insights into how the drug molecules contributed deep into the conduction and valence band edges. The optical responses of anti-TB drugs adsorbed in 2D WSe$_{2}$ and WS$_{2}$ were similar to pristine 2D WSe$_{2}$ and WS$_{2}$. We demonstrated the temperature-dependent release mechanism of our 2D WSe$_{2}$ and WS$_{2}$ drug complexes, confirming the feasibility of releasing the discussed anti-tuberculosis drugs by generating heat through photothermal therapy. These findings hold significant promise for developing innovative drug delivery systems that have enhanced efficacy for targeted and low-toxic TB treatment. } 

\end{tabular}

 \end{@twocolumnfalse} \vspace{0.6cm}
]

\renewcommand*\rmdefault{bch}\normalfont\upshape
\rmfamily
\section*{}
\vspace{-1cm}


\footnotetext{\textit{$^{a}$~Department of Electrical and Electronic Engineering, Bangladesh University of Engineering and Technology, Dhaka 1205, Bangladesh. Tel: +8801300574069; E-mail: ahmedzubair@eee.buet.ac.bd}}

\footnotetext{\dag~Electronic Supplementary Information (ESI) available: [details of any supplementary information available should be included here]. See DOI: 00.0000/00000000.}
\footnotetext{\ddag~Corresponding Author}
\footnotetext{$^{\ast}$ These authors contributed equally in calculations. }





\section{Introduction}
TB is an infectious fatal disease that is one of the deadliest diseases worldwide. A bacterium called \textit{Mycobacterium tuberculosis} primarily attacks the lungs and causes this infection\,\cite{delogu2013biology}. Despite antibiotics being developed, an estimated 10.6 million people are infected by tuberculosis annually, and 1.3 million people died of TB in 2022, according to the World Health Organization (WHO)\,\cite{who}. One of the most prominent reasons for such fatality of this disease is the rise of drug-resistant variants of the \textit{Mycobacterium tuberculosis} \,\cite{almeida2011molecular}. Conventional drug delivery system is worsening the problem as it has a low efficiency and poor bioavailability\,\cite{andresen2005advanced,laffleur2020advances}. Since the traditional drug delivery system is non-controlled and non-specific, even healthy tissues are affected by the drug dosages, and the targeted infected site receives low concentrations of drug dosage. High-concentration dosage in the form of tablets or capsules fails to fix the problem because it aggravates the side effects by harming healthy cells\,\cite{schaberg1996risk}. To combat the issue, it is necessary to develop novel and effective drug delivery systems with enhanced medicinal profiles and therapeutic agent efficacy. Nanomaterial-based targeted drug delivery systems have gained popularity due to their targeted attack on infected cells while not endangering normal tissues and cells of the body\,\cite{jacob2018biopolymer}.

Nanotechnology advancements have revolutionized conventional therapies by enhancing the efficacy of drug delivery strategies\,\cite{dhankhar2010advances}. Nanomaterial-based drug delivery systems have enhanced targeting because their small size, ranging from 1-100 nm, allows them to pass through biological barriers, such as the small intestine and skin, for more effective delivery \cite{cheng2017development}. On top of that,
depending on the drug carrier material, the crystallinity of the drug carrier, and the type of adsorption (chemisorption or physisorption) for different drugs, these nanocarrier/drug complexes release drugs through various mechanisms that ensure a controlled releasing mechanism. Multiple studies have shown that pH, temperature, and chemical reaction-controlled release are common mechanisms for explaining the release of pharmaceuticals from nanocarriers \cite{alvarez2014smart,fan2012controlled}. From previous studies, different nanomaterial-based targeted drug delivery system was reported, including silver (Ag) nanoparticles\,\cite{prasher2020emerging}, Janus and dendrimer particles\,\cite{percec2010self} and quantum dot\,\cite{matea2017quantum}. Different experimental procedures and clinical experiments on animals are occurring regarding nanomaterial-based drug delivery systems.

The investigation of two-dimensional (2D) nanomaterials in the field of therapies and innovative biomedicine has garnered significant attention in recent times, as they provide efficient techniques for disease diagnosis and drug targeting treatment\,\cite{qiu2018omnipotent,ji2019physically}. Graphene and its oxides\,\cite{liu2013graphene,song2020biomedical}, hexagonal boron nitride\,\cite{sukhorukova2015boron} and phosphorene\,\cite{liang2020theoretical} were reported to be useful in targeted drug therapy. 2D materials demonstrate an impressive capability in drug delivery with numerous benefits. The use of 2D materials combined with specific pharmaceutical agents exhibited enhanced efficacy in comparison to alternative nanocarriers, such as nanoparticles, nanotubes, and nanowires, because of their lamella structure, which causes the high surface area-to-mass ratio and other unique physicochemical properties \cite{shi2017cancer,peng2018monolayer,wang2019emerging}. The vast surface area ensured highly efficient drug loading\,\cite{zhang2020recent}. On top of that, 2D nanocarriers have good physical interactions with specific drugs, confirming their proper drug-carrying capability. Transition metal dichalcogenides (TMDs), a group of 2D materials, are very promising in biomedical applications\,\cite{anju2021biomedical,zhu2017design} along with their high utilization in spintronics\,\cite{hoque2022first}, optoelectrocics\,\cite{ifti2020effect}. Moreover, TMD materials' superior light and heat conversion efficiencies make them suitable for photo/thermal-induced tumor photothermal and photodynamic therapy\,\cite{chen20182d}.

Many targeted drug delivery systems were studied for cancer disease, whereas their use in TB treatment received recent attention. To combat TB, some antibiotics like isoniazid (INH), pyrazinamide (PZA) and rifampicin (RIF) are used. Since conventional methods fail to treat this disease, targeted drug delivery systems through nanomaterials have attracted huge interest. Nanoparticles like gold (Au)\,\cite{ali2016gold} and carbon nanotubes\,\cite{jain2013targeted} were used to ensure anti-Tb drugs successfully reach infected lung cells and release the drug. However, 2D materials, including TMDs, will increase efficiency as they have larger planar and good loading and releasing capability. TMDs, such as WSe${_2}$ and WS${_2}$, are widely used in electronics and optoelectronics applications. The monolayers of these materials have direct bandgaps that pave the way for photothermal therapy. WSe${_2}$ and WS${_2}$ are biodegradable and have lower toxicity than graphene and its derivatives\,\cite{appel2016low,teo2014cytotoxicity}. These make them good candidates for targeted drug delivery systems. Exploring the suitability of different highly efficient, nontoxic 2D TMD materials in this field is vital. Moreover, TMDs, such as WSe${_2}$ and WS${_2}$, were not explored as drug delivery systems.

This study investigated the interaction between two prominent anti-TB drugs, INH and PZA, with two promising TMD materials, WSe${_2}$ and WS${_2}$, and the possible adsorption of the drugs on these 2D materials. We calculated various orientations of the adsorbed anti-TB drugs to evaluate the thermal stability of drug/2D TMD complexes from the perspective of adsorption energy. Furthermore, we calculated the band structures of pristine materials and the band structures when drugs were adsorbed on their planer surface. This investigation provided insight into their crystal-based interaction. For further comprehension, projected band structures were calculated to know the orbital contributions of various bands. The density of states of drug atoms on adsorbed complexes was calculated to confirm their interaction with 2D materials. Additionally, we calculated the Mulliken charge analysis and electron density difference to know the nature of the bond and interaction of drug atoms and WS${_2}$, WSe${_2}$. Optical properties were investigated to see the absorption behavior of drug/2D TMD complexes and to check the suitability of using photothermal therapy. Finally, to investigate the controlled releasing criteria, we calculated the change of adsorption energy with respect to the increasing temperature. This study will facilitate designing a nanosystem for targeted and selective anti-TB drug delivery via WS${_2}$ and WSe${_2}$ monolayers.  

\section{Methods}
We performed first-principles calculations using density functional theory\,\cite{engel2011density}, employing generalized gradient approximation (GGA) with Perdew-Burke-Ernzerhof (PBE) parametrization functional for exchange-correlation interactions and Grimme's DFT-D scheme for van der Waals interactions. 
At first, we cleaved a monolayer of tungsten diselenide (WSe${_2}$) and tungsten disulfide (WS${_2}$) from bulk hexagonal symmetry (2-H) crystals. Each unit cell consists of three atoms: one tungsten (W) and two selenium (Se) or sulfur (S) atoms. To prevent interlayer interactions, we introduced a 20 \AA\, vacuum. A (4$\times$4$\times$1) supercell, consisting of a total of 48 atoms, was constructed to serve as the nanocarrier for anti-TB drugs such as INH and PZA. The chemical formulas of INH and PZA are C$_{5}$N$_{3}$H$_{5}$O, C$_{5}$N$_{3}$H$_{5}$O, respectively. We determined the suitable location and orientation for drug adsorption using the Adsorption Locator module\,\cite{akkermans2013monte} in Materials Studio. We selected the most energetically stable orientation and location of drugs for subsequent relaxation and property calculations. The relaxation process was performed using the CASTEP\,\cite{clark2005first} module in Materials Studio, within kinetic energy cutoff of 450 eV, a Gaussian smearing of 0.04 eV, and a (3$\times$3$\times$1) $\Gamma$-centered k-point grid in the Monkhorst-Pack scheme. The iterative procedure was repeated until the convergence threshold of the total energy reached below 10$^{-5}$ eV per atom, the Hellmann-Feynman force among the atoms was below 0.02 eV/\AA, and the stress was below 0.1 GPa. For electronic and optical property calculations, we increased the sampling in the Monkhorst-Pack k-point grid to (5$\times$5$\times$1) $\Gamma$-centered k-points for higher accuracy.
We used the DMol3\,\cite{delley2000molecules} module in Materials Studio to determine the adsorption energies ($E_{ads}$) of the anti-TB drug/WSe${_2}$ (WS${_2}$) complexes. The adsorption energy was calculated by deducting the energies of the drug molecules and the monolayer WSe${_2}$ (WS${_2}$) from the energy of the complex, as given by,
\begin{equation}
E_{ads} = E_{drug/2D TMD} - E_{2D TMD} - E_{drug}
\end{equation}
where $E_{2D TMD}$, $E_{drug}$, and $E_{drug/2D TMD}$ represent the DFT calculated energies of the pure 2D TMDs, the adsorbed drug molecules, and the adsorbed molecule and 2D TMD drug complexes, respectively. Energetically stable configurations were identified by negative values of $E_{ads}$, indicating strong exothermic interactions and thermally stable adsorption. The magnitude of E$_{ads}$ quantified the strength of the interaction.

Electronic band dispersion calculations were performed along the $\Gamma$ (0 0 0) → M (0.0 0.5 0.0) → K (0.333 0.667 0.0) → $\Gamma$ (0 0 0) path for energetically stable configurations. The density of states (DOS) calculations were conducted for pure WSe${_2}$, pure WS${_2}$, drug/ monolayer WSe${_2}$, and drug/ monolayer WSe${_2}$. The projection of DOS on various orbitals of all constituent atoms was performed for drug complexes.

To further understand the nature of adsorption, we conducted a Mulliken charge analysis\,\cite{mulliken1955electronic} and electron density difference (EDD) calculations in CASTEP. Additionally, we calculated the optical properties of the system in CASTEP to analyze the effect of drug adsorption on the 2D sheets. Details of the optical property calculations were provided in \textcolor{blue}{Supplementary Information}. Furthermore, we performed orbital projected band structure calculations using the Quantum Espresso simulation package\,\cite{giannozzi2009quantum,giannozzi2017advanced} to determine the suborbital contributions to the various bands. Moreover, to show the temperature-dependent releasing of the drugs, we used the Phonopy open source package\,\cite{togo2015first} to determine the temperature-dependent portion of Helmholtz free energy. Through Phonopy, supercells with different displacements of atoms were created. After the calculation of the self-consistent field (SCF) of each, the force set was calculated, and through it, thermodynamic properties, including Helmholtz free energy and temperature-dependent vibrational energy, were estimated.

\section{Results}\label{sec2}
\subsection{Geometry structure and adsorption energy for anti-TB drug/2D TMD complexes }\label{subsec2}
\begin{figure*} [htbp]
    \centering
    \includegraphics[scale= 0.4 ]{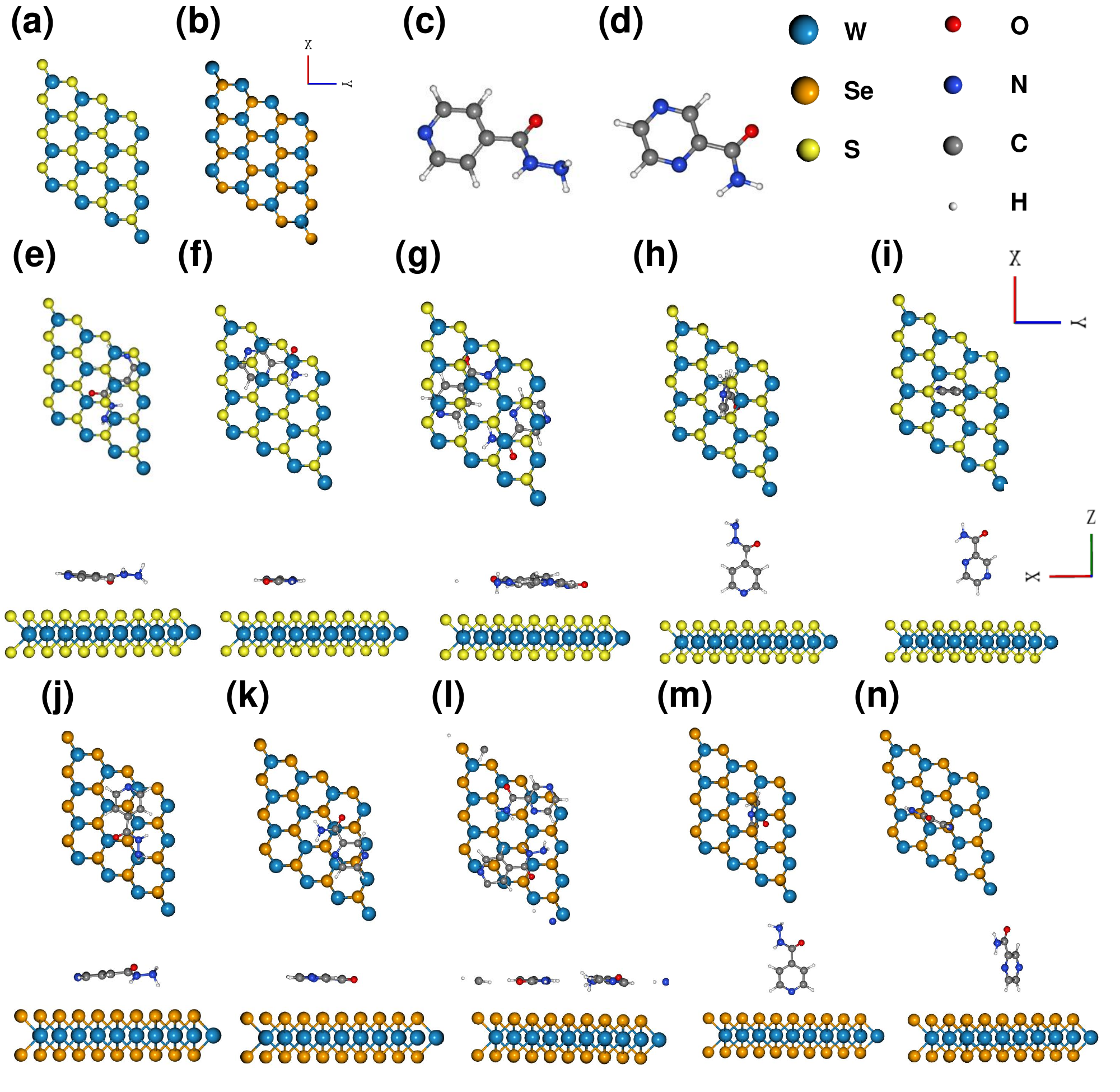}
    \caption{Optimized structures of (a) WS$_{2}$, (b) WSe$_{2}$, (c) INH, (d) PZA. Top and side views of the optimized TMD complexes of (e) INH(H)/WS$_{2}$, (f) PZA(H)/WS$_{2}$, (g) INH/PZA(H)/WS$_{2}$, (h) INH(V)/WS$_{2}$, (i) PZA(H)/WS$_{2}$, (j) INH(H)/WSe$_{2}$, (k) PZA(H)/WSe$_{2}$, (l) INH/PZA(H)/WSe$_{2}$, (m) INH(V)/WSe$_{2}$, (n) PZA(V)/WSe$_{2}$. Here,  drug(H) indicates the horizontal position of drug atoms, and drug(V) indicates the vertical position. }

    \label{figure 1}
\end{figure*}

We relaxed pristine structures of WS$_{2}$, WSe$_{2}$, INH, and PZA, which are shown in Figs. \ref{figure 1} (a), (b), (c), (d). INH and PZA drugs have a hexagonal plane and a tilted -CON$_{2}$H$_{3}$ and -CONH$_{2}$ functional groups. The transition metal tungsten (W) is sandwiched between two chalcogen layers of sulfur (S) atoms in 2-H WS$_2$. The bulk WS$_{2}$ consists of many layers of monolayer WS$_2$, and layers are packed due to weak van der Waals forces. A monolayer of WS$_{2}$ can be found through physical or chemical processes. A hexagonal phase is seen in the WS$_{2}$ structure where three S atoms are bonded to a single W atom, and prominent Coulombic interaction is found between S and W molecules. WSe$_{2}$ has a similar geometric structure. Lattice constants, bond length between W and S/Se atoms, and bond angles of two optimized structures were investigated. The results are shown in Table \ref{figure 1}. The bond length of W-S was found to be 2.42\,\AA\, and for W-Se, the length was 2.53 \AA. These results were consistent with the previous studies\,\cite{bo_len,ifti2020effect}.

\begin{table*}[h]
\caption{Calculated Structural Parameters }\label{tab1}%
\begin{tabular*}{1\textwidth}{@{\extracolsep{\fill}}lllllll}

\hline
Crystal structures &lattice constant &bond length &bond length &\angle $S-W-S$ &\angle $Se-W-Se$ \\ 
 &(a=b in \AA) & W-S (\AA) &   W-Se (\AA)  \\ 
 \hline
                 
WS$_{2}$ (4$\times$4$\times$1)  & 12.72   & 2.42  &  &  80.94°  \\
WSe$_{2}$ (4$\times$4$\times$1)  &    13.17   &   & 2.53 & & 81.13° \\
\hline

\end{tabular*}
\end{table*}

Figs. \ref{figure 1} (e)-(i) represents the different adsorption configurations of INH and PZA drugs on monolayer WS$_{2}$ surface. From top and side views, it is demonstrated that drugs were adsorbed in both horizontal and vertical directions. In Fig. \ref{figure 1} (g) it is seen that both drugs were simultaneously adsorbed in the horizontal direction for WS$_{2}$. Here, horizontal configuration implies that the drug and 2D TMDs were in a parallel direction, and vertical configuration means perpendicular orientation of drug structures. Figs. \ref{figure 1} (j)-(n) shows the 2D TMD drug complexes of horizontal and vertical adsorption on monolayer WSe$_{2}$. Table \ref{distance} shows the calculated adsorption energy and vertical distances between anti-TB drugs and 2D TMDs. In 2D TMD drug complexes, the configurations suggest that physisorption occurred rather than chemisorption. For vertical structures of drug/2D TMD, we used the notation of drug(V)/2D TMD, and for horizontal structures, we used d, for all horizontal configurations. Drug (V)/WS$_{2}$ and drug(V)/WSe$_{2}$ had higher adsorption energies than other configurations, making them metastable. The most thermally stable structure was found when both INH and PZA drugs were adsorbed simultaneously. INH/PZA(H)/WSe$_{2}$ configuration was the thermally strongest 2D TMD monolayer complex with an adsorption energy of -2.35 eV. PZA(V)/WS$_{2}$ configuration had the highest adsorption energy of -0.28 eV among the studied configurations. Comparing thermal stability through formation energy of the structures, INH/PZA(H)/WSe$_{2}$ ${>}$ INH/PZA(H)/WS$_{2}$ ${>}$ INH(H)/WSe$_{2}$ ${>}$ PZA(H)/WS$_{2}$ ${>}$ INH(H)/WS$_{2}$ ${>}$ PZA(H)/WS$_{2}$INH(V)/WSe$_{2}$ ${>}$PZA(V)/WS$_{2}$ ${>}$ INH(V)/WS$_{2}$ ${>}$PZA(V)/WS$_{2}$. If the adsorption energy of a drug/2D TMD complex gets much lower, then stability may be higher but it will be difficult to release the drug in a convenient way. So, adsorption energy must be in the moderate range for drug release and structural stability. All the drug(V)/2D TMD complexes had much higher adsorption energy so they might be in a metastable state where their uncontrolled release characteristic is more probable. As a result, we emphasized drug(H)/2D TMD complexes in this paper.
\begin{table}[h]
\caption{Calculated vertical adsorbed distance,d (in \AA), adsorption energy, E$_{ads}$ (in eV unit) of INH, PZA drugs on WS$_2$ and WSe$_2$ surfaces } \label{distance}%
\begin{tabular*}{0.48\textwidth}{@{}ccccc@{}}
\hline

\centering

\multirow {3}{*} {Drug structures} & \multicolumn{2}{c}{WS$_2$} & \multicolumn{2}{c}{WSe$_2$}  \\  
                  &  d & E$_{ads}$ &  d & E$_{ads}$ \\ 
                  & (\AA) & (eV) &(\AA) & (eV) \\ 
                  \hline
INH (H)    &3.46   &-0.96  &3.89 &-1.25  \\
PZA (H)  &3.38   &-0.99  &3.59 &-1.21  \\
INH/PZA (H)  &3.11   &-1.97  &3.16 &-2.35  \\
INH  (V)  &3.32   &-0.48 &2.81 &-0.80  \\
PZA (V)   &3.49   &-0.28  &2.81 &-0.74  \\
\hline

\end{tabular*}
\end{table}

\subsection{ Electronic Properties }\label{subsec2}
\subsubsection{WS$_{2}$ and WSe$_{2}$ Monolayers}\label{subsubsec2}
\begin{figure*}[htbp]
    \centering
    \includegraphics[scale= 0.4 ]{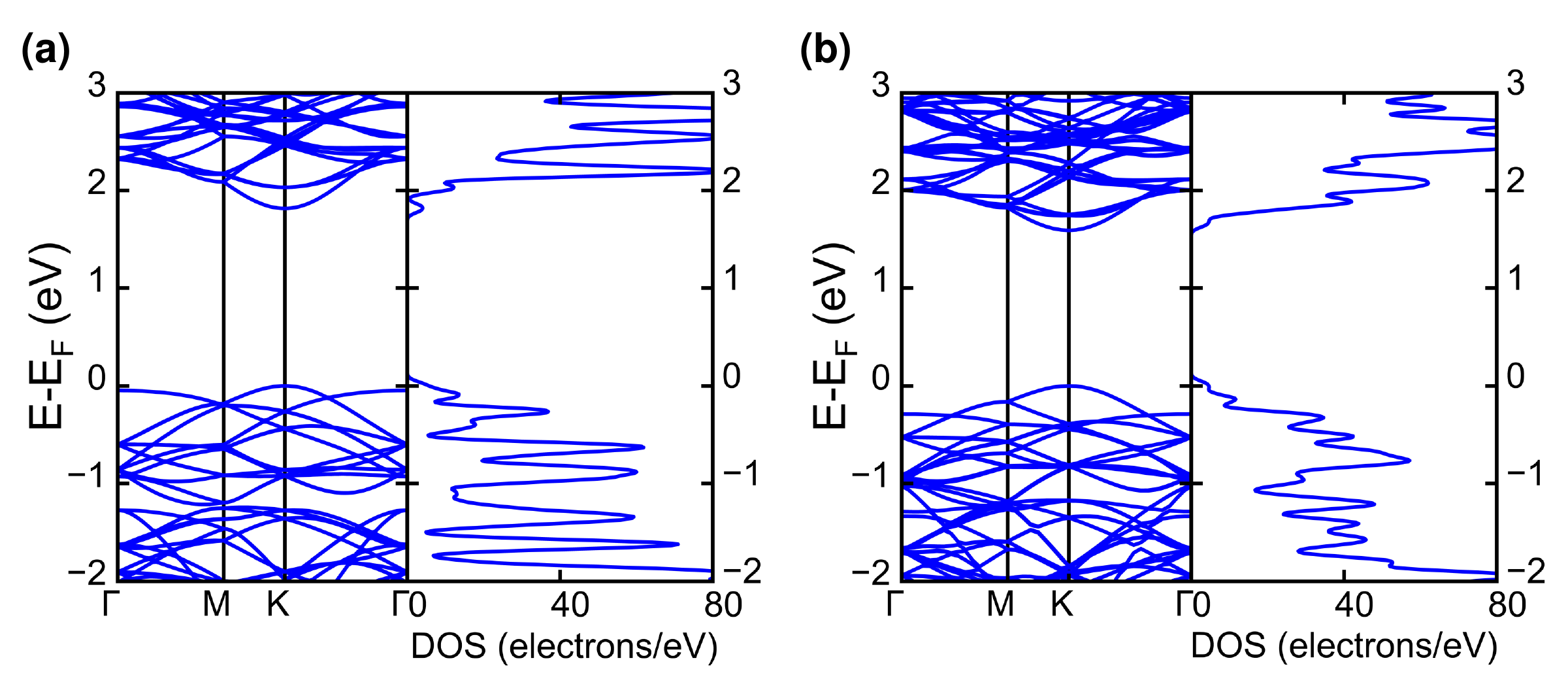}
    \caption{Calculated electronic band structures and corresponding DOS of pristine (a) WS$_{2}$ and (b) WSe$_{2}$ monolayers.}
    \label{pristine_band}
\end{figure*}
We calculated and analyzed the band structures and DOS of horizontally adsorbed drugs on 2D nanocarriers. At first, the band structure and DOS of the monolayer supercells of WS$_{2}$ and WSe$_{2}$ were calculated. Figs. \ref{pristine_band} (a) and (b) show the band structure and DOS of pristine WS$_{2}$  and WSe$_{2}$. The bandgap of pristine WS$_{2}$ was 1.81 eV, and the bandgap of WSe$_{2}$ was 1.60 eV, and the results matched with previously reported values\,\cite{chaurasiya2018strain,ernandes2021indirect,ifti2020effect}. Both of the nanocarriers had direct bandgap, and the bandgap occurred at the high symmetry K point. 

\begin{figure*}[htbp]
    \centering
    \includegraphics[scale= 0.4  ]{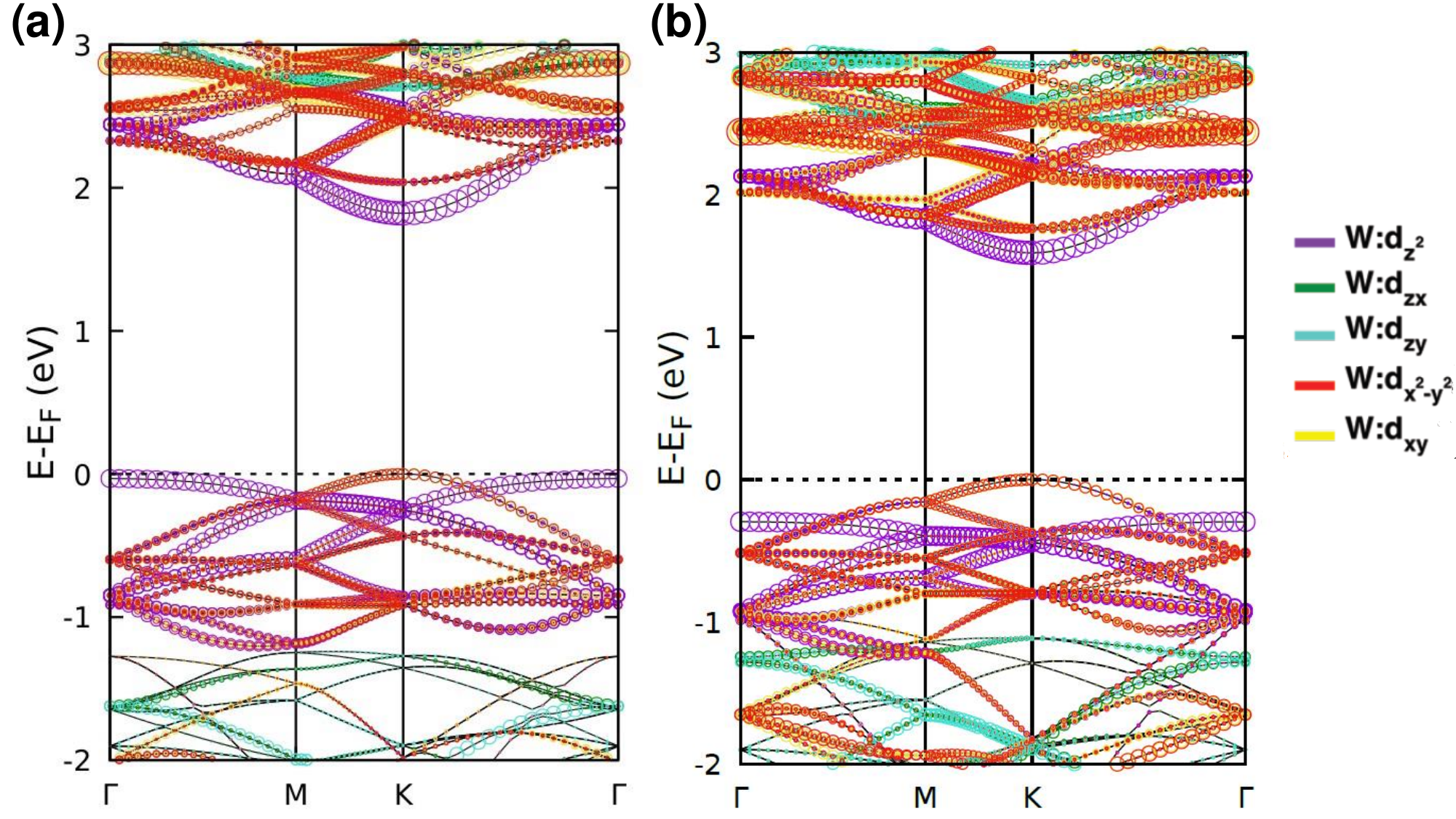}
    
     \caption{Orbital projected band diagrams showing the contribution of d-orbitals of W atom of (a) WSe$_{2}$ and (b) WSe$_{2}$. In the figures, the amount of contributions is indicated by the size of the circles, and corresponding colors illustrate the contributing sub-orbitals.}
    \label{projection_pris}
\end{figure*}

The band projection on atomic orbitals was performed to understand the band structure better, and the contributions of different orbitals were evaluated. The d-orbitals of the transition metal W atom mainly contributed to the valance and conduction band edges. For WS$_{2}$, d\textsubscript{x\textsuperscript{2}-y\textsuperscript{2}} contributed 36.68\% and d\textsubscript{xy} contributed 36.66\% in valance band maxima (VBM). Other main contributions of VBM came from p\textsubscript{x} (10.4\%) and p\textsubscript{y} (10.4\%) orbitals of S atoms. The conduction band minima (CBM) mainly consisted of d\textsubscript{z\textsuperscript{2}} orbital, with a contribution of 80.32\%. The s orbital of the W atom contributed more than 10\% to CBM. For WSe$_{2}$, the involvement of d orbitals with their contribution percentage remained almost the same. The d\textsubscript{x\textsuperscript{2}-y\textsuperscript{2}} contributed 38.59\% and d\textsubscript{xy} contributed 38.11\% in valance band maxima (VBM) of WS$_{2}$, while d\textsubscript{z\textsuperscript{2}} contributed to 78.69\% to CBM. Figs. \ref{projection_pris} (a) and (b) show the orbital projected band diagram for pristine WS$_{2}$ and WSe$_{2}$ monolayers. 

\subsubsection{Drug(H)/2D TMD Complexes}\label{subsubsec2}

\begin{figure*}
\centering
    \includegraphics[scale= 0.3  ]{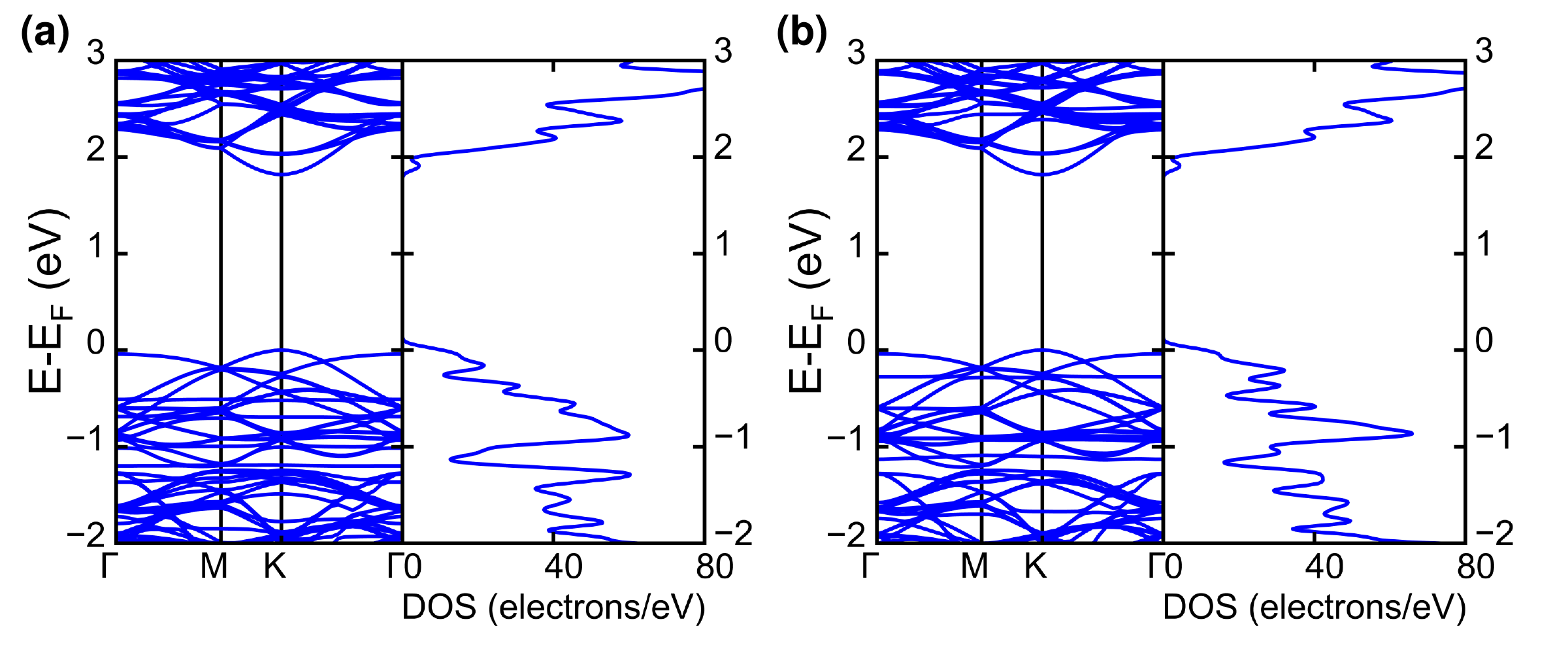}
     \includegraphics[scale= 0.3  ]{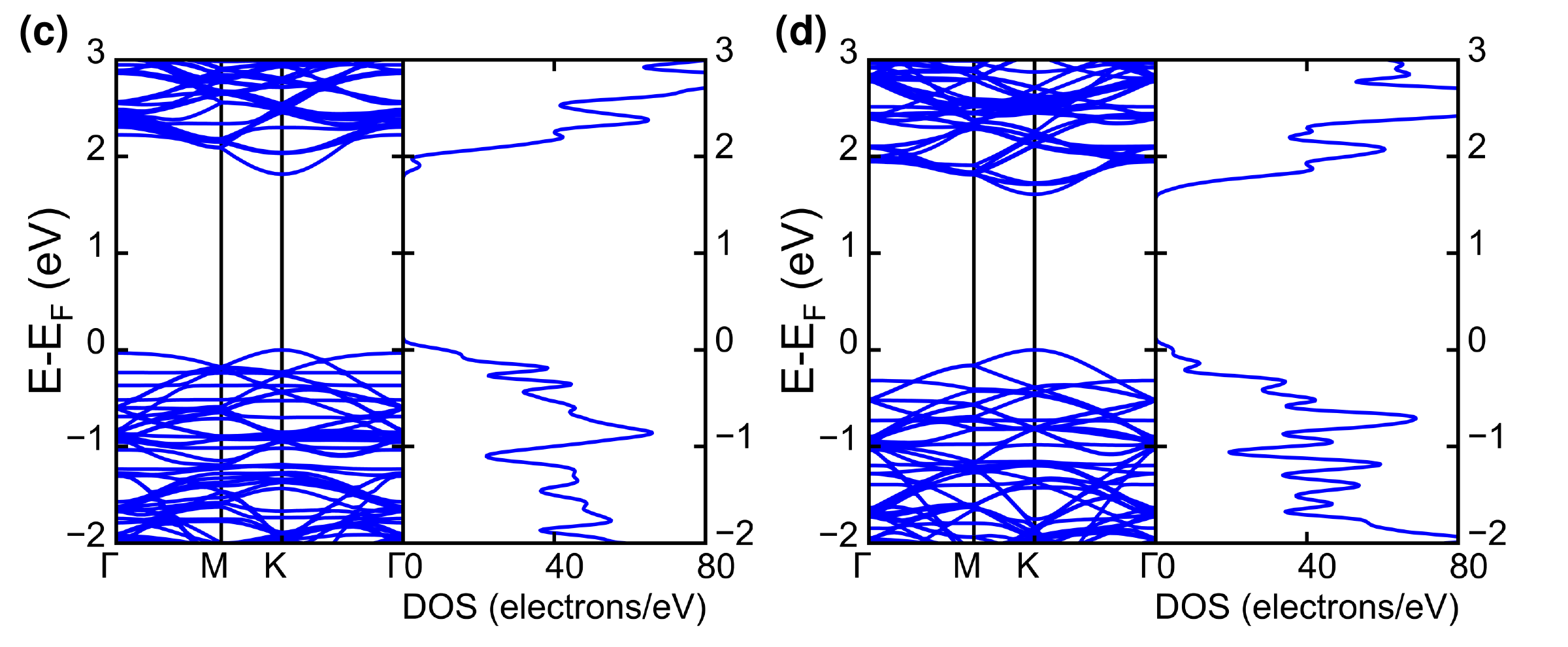}
     \includegraphics[scale= 0.3  ]{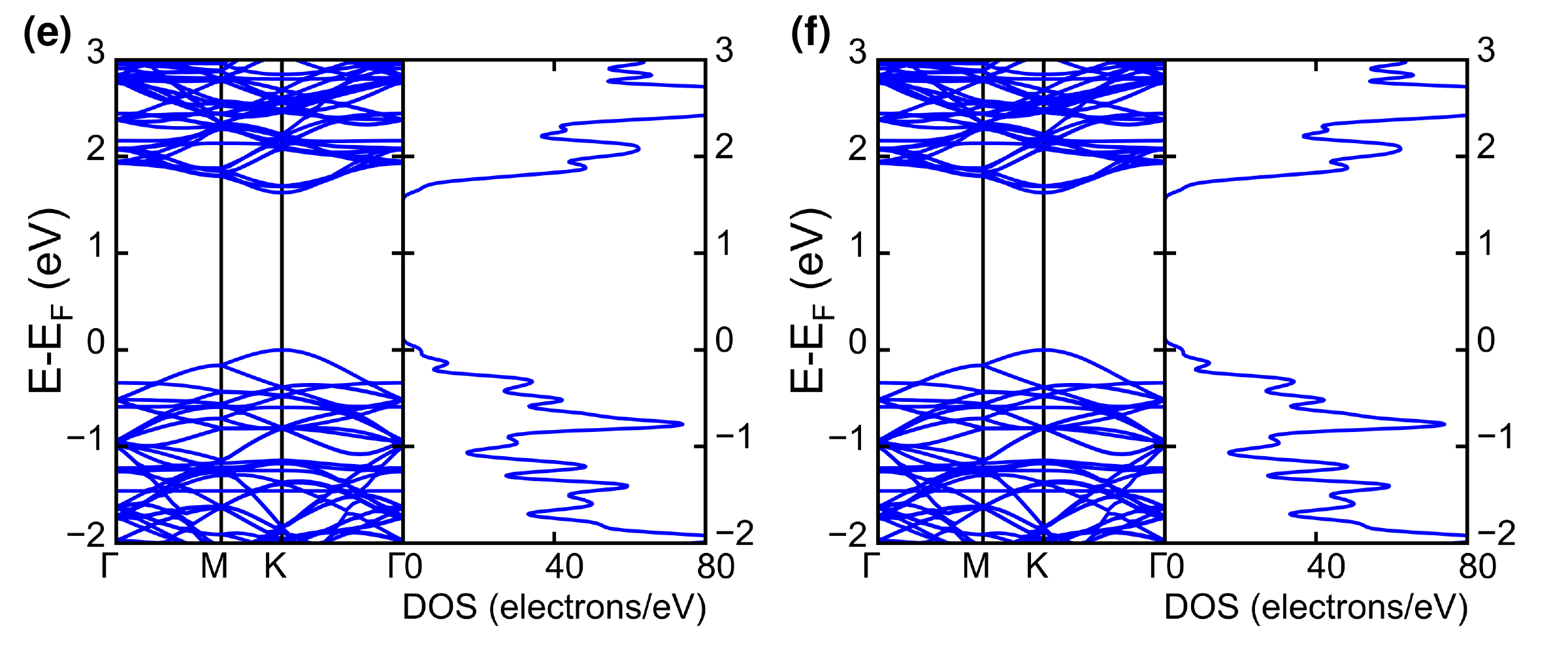}
     \caption{Electronic band structures and DOS of (a) INH(H)/WS$_{2}$, (b) PZA(H)/WS$_{2}$, (c) INH/PZA(H)/WS$_{2}$, (d) INH(H)/WSe$_{2}$, (e) PZA(H)/WSe$_{2}$, (f) INH/PZA(H)/WSe$_{2}$. }
    \label{complex band}
\end{figure*}

For both 2D TMD drug complexes, band structures were dominated by pristine 2D TMD sheets. It was found that the drug atoms did not significantly modify the electronic bands when they were adsorbed on the TMDs. The bandgap of the drug(H)/2D material complexes was almost identical to their pristine material. Besides, the direct bandgap behavior of WS$_2$ and WSe$_2$ remained the same for all complexes. DOS of the structures matched with corresponding band structures. Figs. \ref{complex band} (a)-(c) demonstrates WS$_{2}$ sheet adsorbed by INH, PZA, and a combination of INH/PZA, respectively. Figs. \ref{complex band} (d)-(f) shows the calculated results for drug absorbed WSe$_{2}$ complexes. Band structures and DOS of Vertically adsorbed INH on WS$_2$ and WSe$_2$ are discussed in \textcolor{blue}{Supplementary Information}

\begin{figure*}[htbp]
\centering
   
    \includegraphics[scale= 0.3 ]{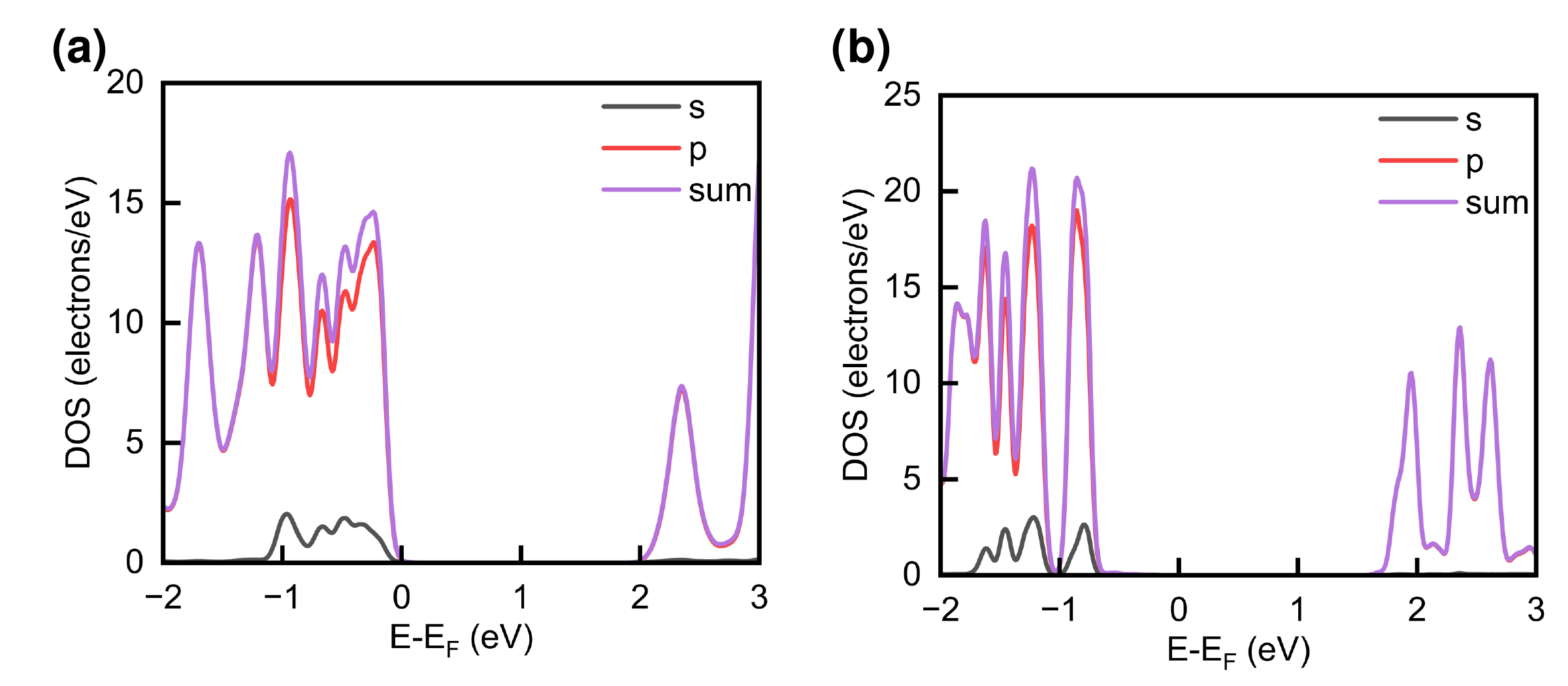}
     \caption{Calculated projected DOS of INH, PZA drug atoms for (a) INH/PZA(H)/WS$_{2}$ and (b) INH/PZA(H)WSe$_{2}$.}
    \label{drug dos}
\end{figure*}

\begin{figure*}[htbp]
    \centering
    \includegraphics[scale= 0.4   ]{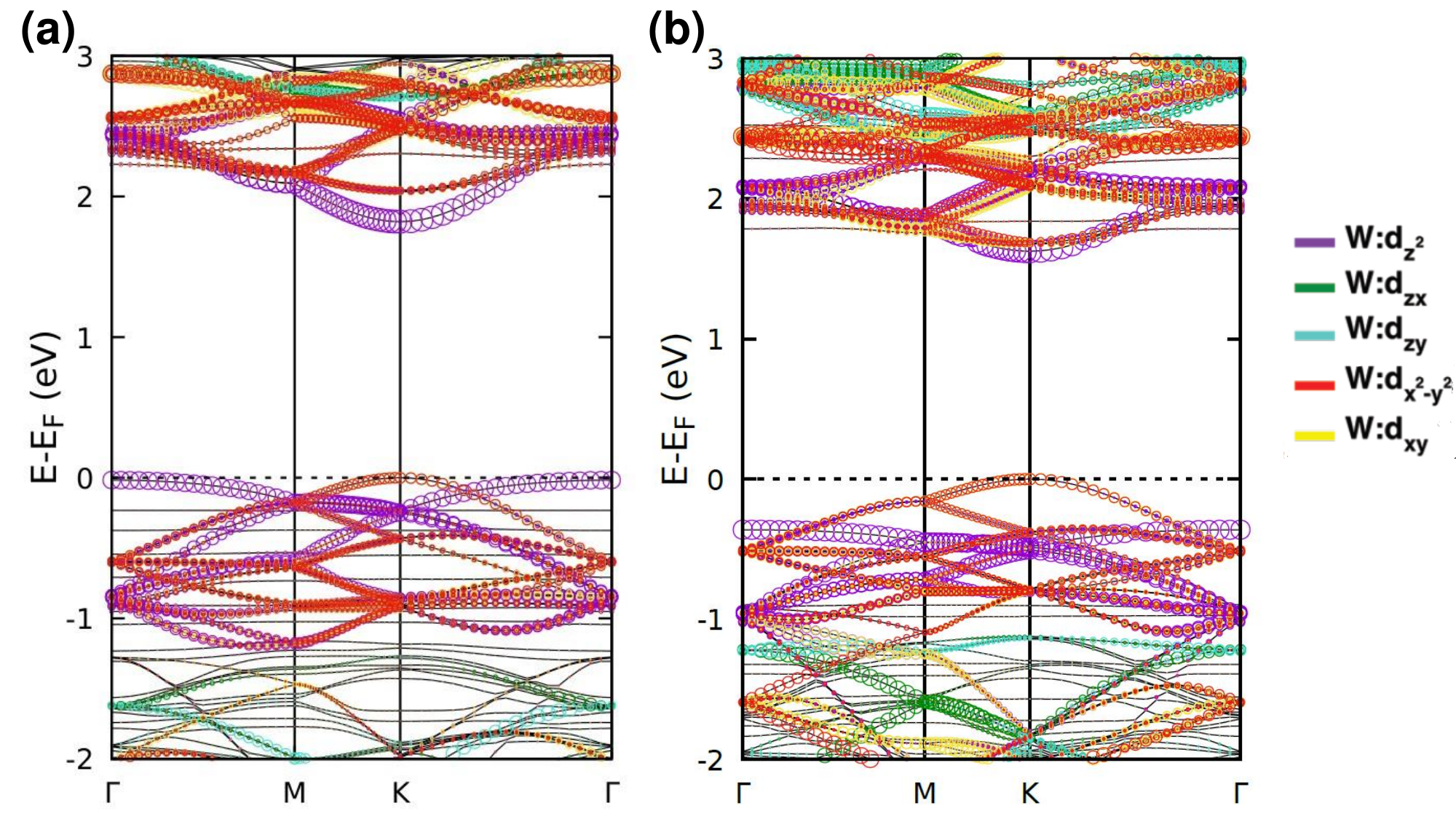}
    
     \caption{Orbital projected band diagrams showing the contribution of d orbitals of W atom for (a) INH/PZA(H)/WS$_{2}$ and (b) INH/PZA(H)/WSe$_{2}$ are illustrated. In the band diagrams, the amount of contributions is indicated by the size of the circles, and corresponding colors illustrate the contributing sub-orbitals. }
    \label{projection}
\end{figure*}

 The projected DOS of drug atoms in drug/2D TMD complexes was calculated for further investigation. Figs. \ref{drug dos} (a) and (b) depict the DOS of drug atoms in the two structures that were energetically most favorable, INH/PZA(H)/WS$_2$ and INH/PZA(H)/WSe$_2$. As new states of drugs were introduced to the complex structures, it is evident that interactions occurred between drug atoms and 2D materials. TB drug atoms that were adsorbed did not add new states to the pristine bandgap region, and that's why pristine structures dominated the bandgap of the 2D TMD monolayer complexes. However, drug atoms have added states below the VBM and above the CBM. Besides, it can be seen that the p orbital of the drug atoms predominantly contributed to the new states where the s orbital had little contribution in valance bands and almost negligible presence in conduction bands. Details of individual drug atoms' contribution in DOS are discussed in \textcolor{blue}{Supplementary Information}. Furthermore, we calculated the projected band structure for those two energetically most stable structures and found that the contributions of d orbitals are almost similar to their pristine structures. For INH/PZA(H)/WSe$_2$, d\textsubscript{x\textsuperscript{2}-y\textsuperscript{2}} contributed 39.65\% and d\textsubscript{xy} contributed 39.65\% in VBM where CBM maninly consisted of d\textsubscript{z\textsuperscript{2}} orbital, with the contribution of 80.55\%. However, insignificant changes in d suborbital contributions were observed in INH/PZA(H)/WS$_2$ compared to those of pristine WS$_2$, as seen in Fig.\,\ref{projection}.

\subsection{ Charge Density Analysis }\label{subsec2}
After confirming the energy stability of drug adsorption, comprehensive analyses were conducted using Mulliken charge analysis and electron density difference (EDD) to elucidate the nature and mechanisms of drug adsorption on monolayer WSe$_{2}$ and WS$_{2}$. Notably, no covalent bonds were formed between the drug molecules and the 2D TMD materials, and the adsorption distances fell within the optimal range of physisorption. We determined the electron density difference, $\Delta n_{drug/2D TMD}$, of the 2D TMD complexes before and after the drug molecules' physisorption using the formula below to explore the interaction between molecules and 2D TMDs.

\begin{equation}
\Delta n_{drug/2D TMD} = n_{drug/2D TMD}-n_{2D TMD}-n_{drug}
\end{equation}

Mulliken charge transfer analysis revealed that the average charge per atom exhibited slight changes following adsorption. Electrons were transferred from the drug molecules to the drug(H)/2D TMD complex interface, specifically to the electrophilic groups (CON$_{2}$H$_{3}$ in INH and CONH$_{2}$ in PZA) of the drug molecules. The electrophilic groups of isoniazid (INH) exhibited an attractive force towards the electrons present in the 2D TMDs. This interaction led to the polarization of the complex situated underneath the adsorption site, thereby inducing intramolecular charge transfer. The transfer of electrons between the drug molecules and the WSe$_2$ and WS$_2$ surfaces was not uniform. The atoms in the drug molecules that were closer to the interface experienced the greatest electron transfer, while the atoms that were further away experienced negligible electron transfer. The average charge density difference for each drug molecule is summarized in Tables \ref{tabb1} and \ref{tabb2}.  
\begin{table}[h]
\caption{Average Mulliken charge of atoms of WSe$_2$ drug complex}\label{tabb1}%
\begin{tabular*}{0.48\textwidth}{@{}lllll@{}}
\hline

Atoms & WSe$_2$  & INH/WSe$_2$ & PZA/WSe$_2$ & INH/PZA/WSe$_2$ \\  
                  &  ($|e|$) &  ($|e|$) &  ($|e|$) &  ($|e|$) \\
                  \hline

H    & 0.378   & 0.266  & 0.292 &  0.283  \\
C    &    0   & -0.068  & 0.034 & -0.023  \\
N    & -0.565   & -0.510  & -0.520 & -0.517  \\
O    & -0.575   & -0.530  & -0.550 & -0.535  \\
Se    & 0.100   & 0.121  & 0.113 &  0.139  \\
W    & -0.190   & -0.203  & -0.196 & -0.212  \\
\hline

\end{tabular*}
\end{table}

\begin{table}[h]
\caption{Average Mulliken charge of atoms of WS$_2$ drug complex}\label{tabb2}%
\begin{tabular*}{0.48\textwidth}{@{}lllll@{}}
\hline

Atoms & WS$_2$  & INH/WS$_2$ & PZA/WS$_2$ & INH/PZA/WS$_2$ \\  
                  &  ($|e|$) &  ($|e|$) &  ($|e|$) &  ($|e|$) \\ 
                  \hline

H    & 0.171   & 0.356  & 0.380 &  0.356  \\
C    & 0.029   & 0.061  & 0.048 & -0.003  \\
N   & -0.196  &-0.540  &-0.546 & -0.540  \\
O   & -0.372   &-0.560  &-0.590 & -0.565  \\
S   & -0.030   &-0.028  &-0.028 & -0.024  \\
W   &  0.060   & 0.059  & 0.058 &  0.059  \\
\hline

\end{tabular*}
\end{table}

Drug molecules inside the INH(H)/WSe$_2$ complex demonstrated the most significant gain in electron density, with the C, N, and O atoms gaining 0.068e, 0.055e, and 0.045e, respectively. Conversely, the H atoms in these molecules lost 0.112e.
The molecules of the drug within the PZA(H)/WSe$_2$ complex also gained electrons, but the gains were smaller than those observed in the INH(H)/WSe$_2$ complex (0.045e, 0.034e, and 0.025e for N, C, and O atoms, respectively). The H atoms in these molecules also lost electrons, but the losses were also smaller than those observed in the INH(H)/WSe$_2$ complex (0.086e).
The drugs contained within the INH/PZA(H)/WSe$_2$ complex showed the greatest gains again, with the N, O, and C atoms gaining 0.048e, 0.040e, and 0.023e, respectively. The H atoms in these molecules also lost electrons, but the losses were similar to those observed in the PZA(H)/WSe$_2$ complex (0.095e).

Drug molecules inside the PZA(H)/WS$_2$ complex gained electrons, with the greatest gains observed for the N and O atoms (0.344e and 0.118e, respectively). Conversely, the H and C atoms in these molecules lost electrons (0.112e, 0.019e).
The molecules of the drug within the INH(H)/WS$_2$ complex also gained electrons, but the gains were higher than those observed in the PZA(H)/WS$_2$ complex (0.188e, 0.344e and 0.089e for O, N and C atoms, respectively). The H atoms in these molecules also lost electrons, but the losses were also higher than those observed in the PZA(H)/WS$_2$ complex (0.185e).
The drugs contained within the INH/PZA(H)/WS$_2$ complex showed the largest gains in electron density among the three, with the greatest gains again observed for the C, O, and N atoms (0.0315e, 0.193e, and 0.344e respectively). The H atoms in these molecules also lost electrons, but the losses were similar to those observed in the PZA(H)/WS$_2$ complex (0.185e).

\begin{figure*}[htbp]
    \centering
    \includegraphics[trim={2cm 0cm 2cm 0cm}, clip,scale= 0.4  ]{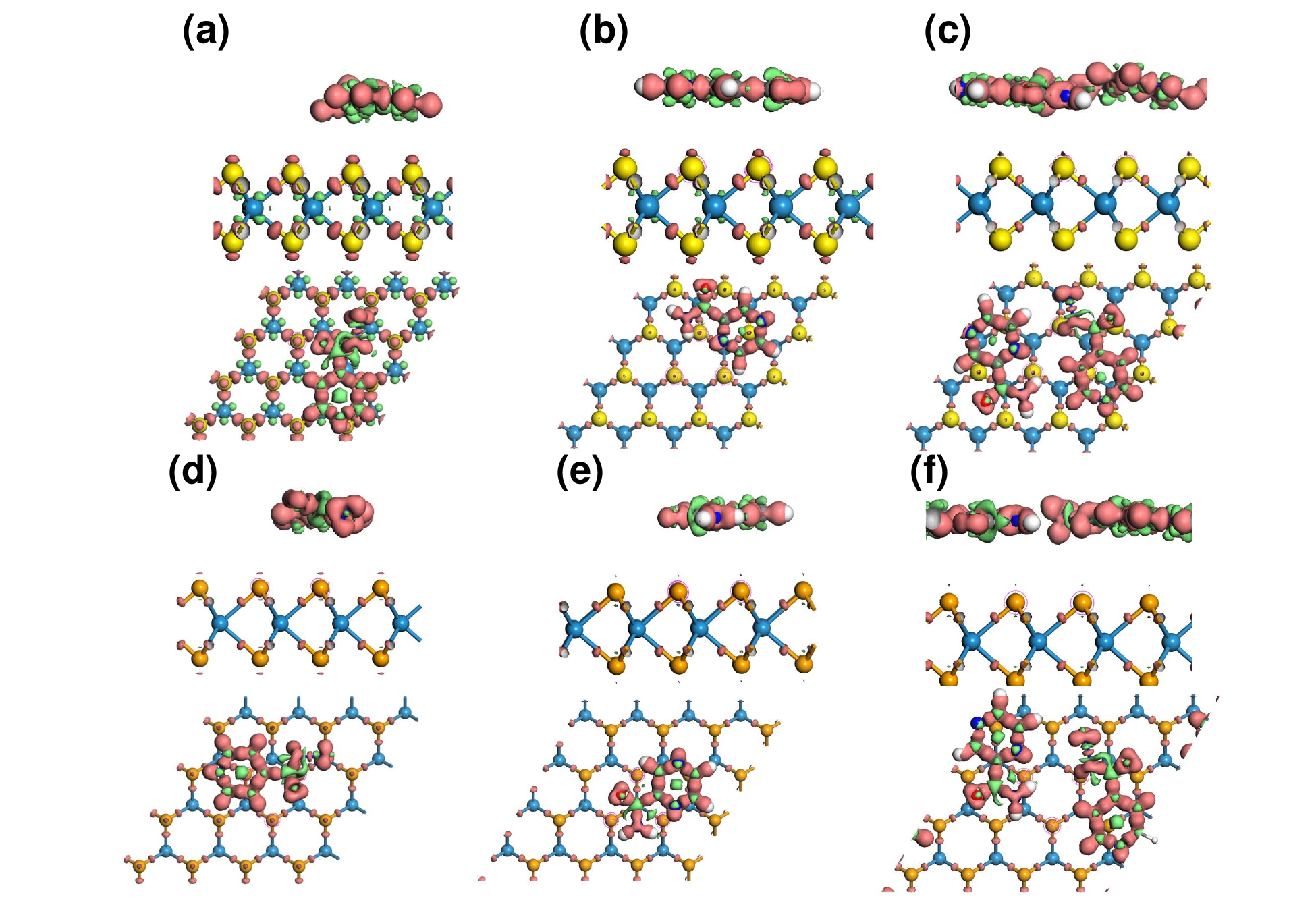}
    \caption{Electron density difference plots of (a) INH(H)/WS$_{2}$, (b) PZA(H)/WS$_{2}$, (c) INH/PZA(H)/WS$_{2}$, (d) INH(H)/WSe$_{2}$, (e) PZA(H)/WSe$_{2}$, (f) INH/PZA(H)/WSe$_{2}$. Red and green colors depict the accumulation and depletion of electron density of the structures. The iso-surface value is 0.1 }
    \label{EDD}
\end{figure*}

Overall, the Mulliken charge transfer analysis results suggest that the drug molecules in all six complexes interacted with the WSe$_2$ and WS$_2$ surfaces similarly, with the greatest electron transfer occurring to the N, C, and O atoms. The H atoms in these molecules lost electrons, but the losses were smaller than those observed for the N, C, and O atoms. This electron sharing was found to be insufficient for the formation of additional covalent bonds but adequate for physisorption. Fig.\,\ref{EDD} visually depicts the accumulation of electrons on the Se and S atoms, while the drug functional groups experienced a depletion of electrons. Here, an iso-surface value of 0.1 was used. Considering the significant adsorption distances and the low magnitude of charge transfer, it can be concluded that the drug molecules undergo physisorption on the WSe$_{2}$ and WS$_{2}$ monolayers.

\subsection{ Optical Property }\label{subsec2}
The WSe$_2$ and WS$_2$ drug complexes can be used in phototherapy. Optical properties of WSe$_2$ and WS$_2$ and their drug complexes were calculated. Absorption coefficient vs wavelength is a critical parameter for nanomaterial-based drug delivery systems. Absorption coefficients are calculated from the imaginary part of the complex dielectric constant of pristine structures and their complexes. (See \textcolor{blue}{Supplementary Information} for the detailed formulation). The relationship between absorption coefficient, $\alpha$, and dielectric constants, $\varepsilon_{1}$ and $\varepsilon_{2}$ is given by the following equation.
\begin{equation}
\label{opticaleqn}
\alpha(\omega)=\frac{4 \pi \kappa(\omega)}{\lambda}=\frac{4 \pi}{\lambda \sqrt{2}}\left(\sqrt{\varepsilon_{1}^{2}(\omega)+\varepsilon_{2}^{2}(\omega)}-\varepsilon_{1}(\omega)\right)^{1 / 2},
\end{equation}
Here, $\kappa$ is the extinction coefficient and $\lambda$ is the wavelength

\begin{figure*}
    \centering
    \includegraphics[scale= 0.4 ]{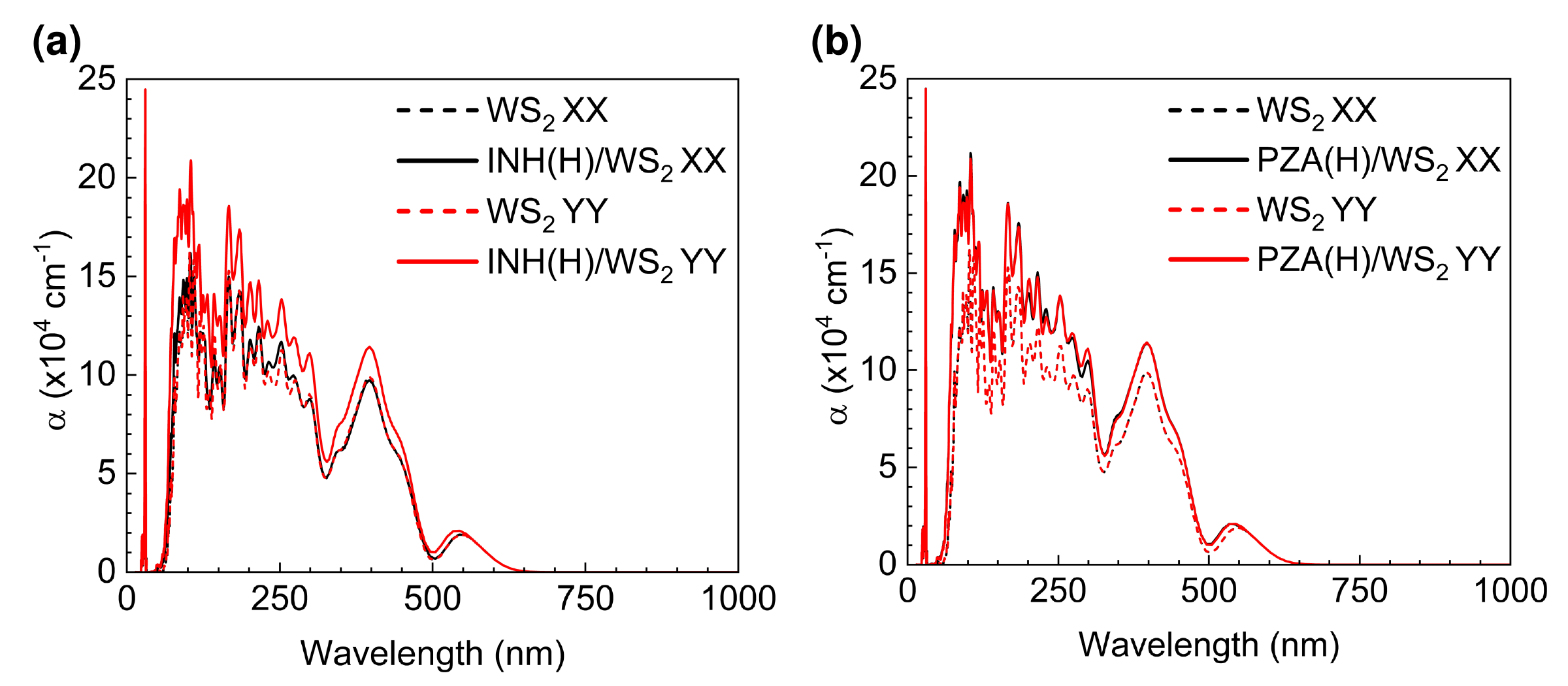}
     \includegraphics[scale= 0.4 ]{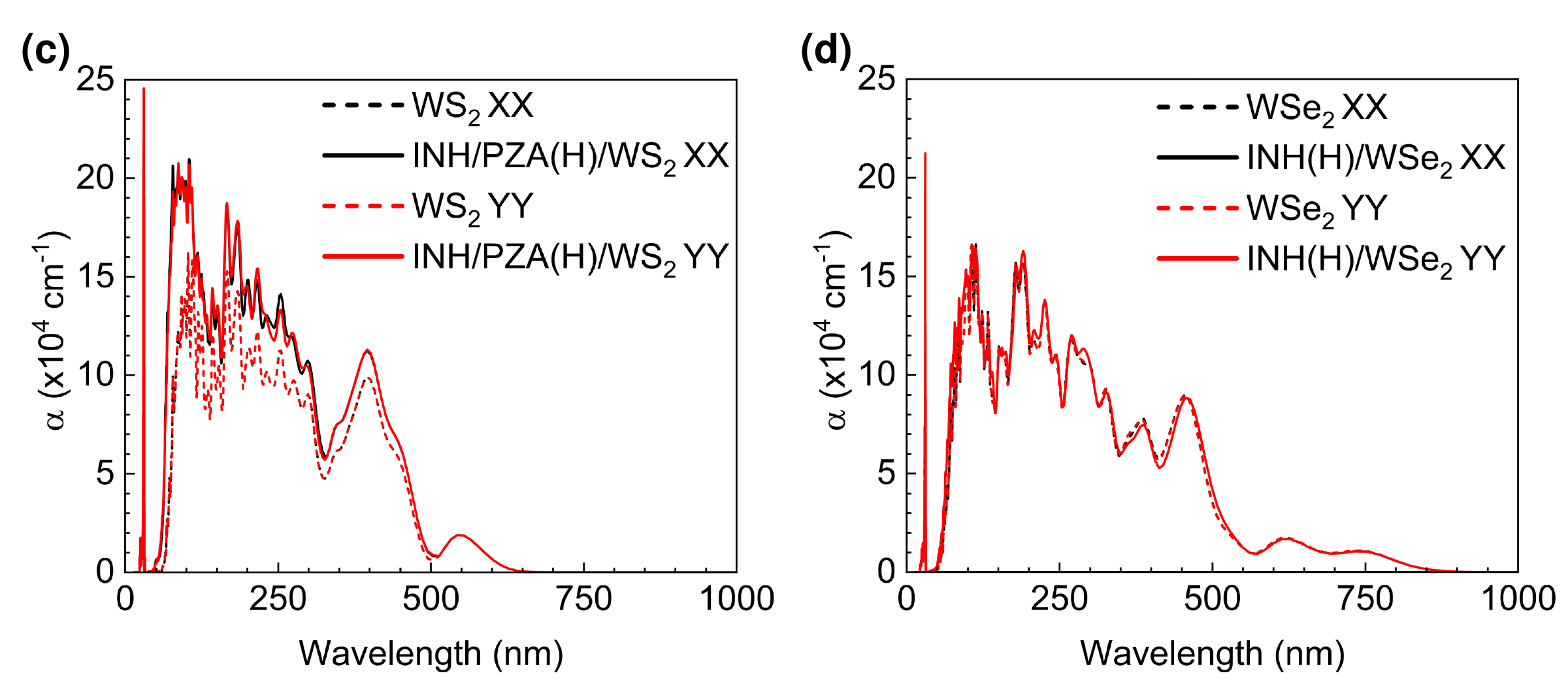}
     \includegraphics[scale= 0.4]{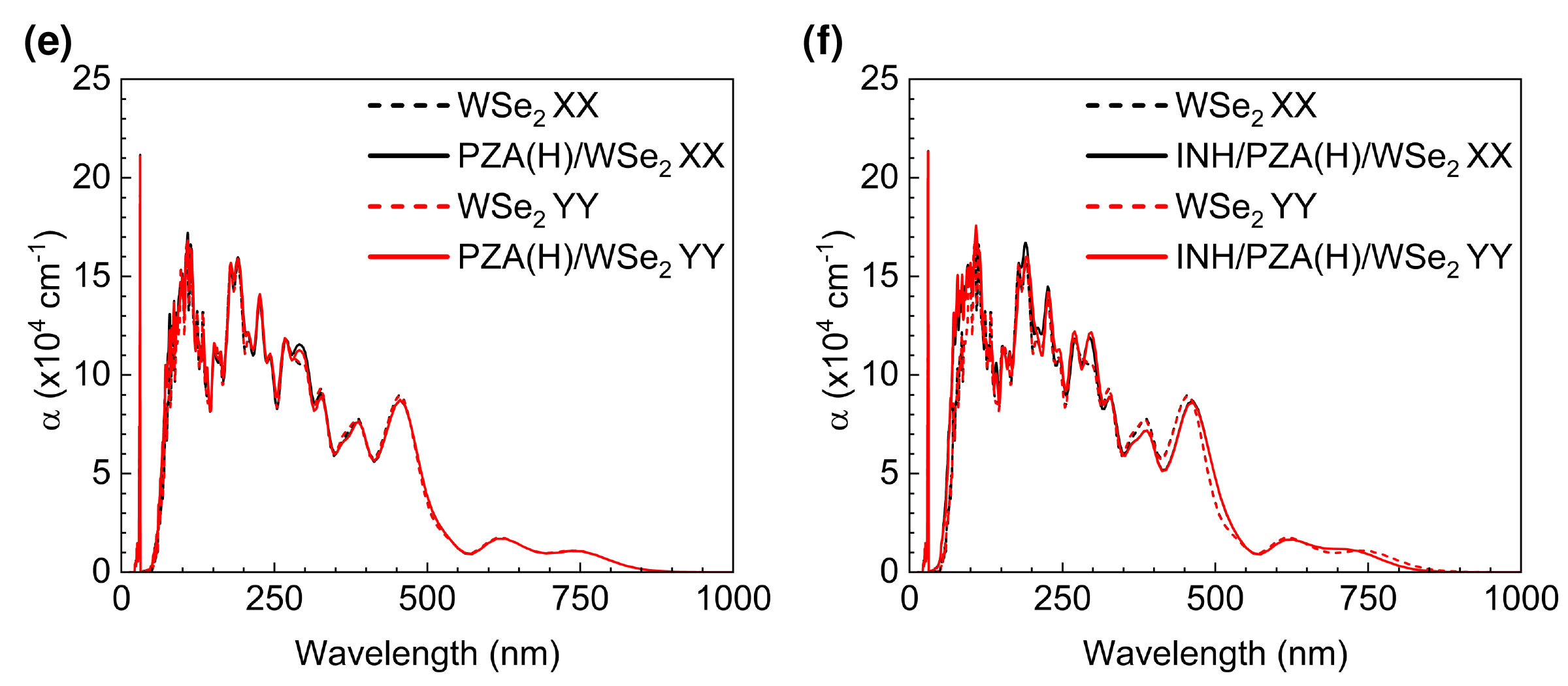}
     \caption{Absorption coefficient, $\alpha$ vs wavelength figure for (a) INH(H)/WS$_{2}$, (b) PZA(H)/WS$_{2}$, (c) INH/PZA(H)/WS$_{2}$,(d) INH(H)/WSe$_{2}$, (e) PZA(H)/WSe$_{2}$, (f) INH/PZA(H)/WSe$_{2}$; in every figure corresponding pristine WS$_{2}$ or WSe$_{2}$ trend is denoted by dotted line to visualize the comparison between graphs. }
    \label{optical}
\end{figure*}

For the drug complexes to be utilized in photothermal therapy, there should be good absorption within the visible range (wavelengths of 400–700 nm) or near-infrared (NIR) range (700–1,350 nm)\,\cite{PTT}. It is because we cannot use light of all wavelengths on the body as it will damage the tissues and cells of the body. Figs.\,\ref{optical} (a)-(f) show the optical behavior of drug(H)/WS$_2$ and drug(H)/WSe$_2$ complexes compared with the pristine behavior. It is evident that a good absorption peak was found in both the visible range and the NIR range for drug (H)/WSe$_2$ structures. However, for drug(H)/WS$_2$ structures, a notable peak is found only within the visible light range. The PZA and INH structures have a noteworthy absorption peak between 3.5 and 4.0 eV \cite{liang2020theoretical}. When they were adsorbed by WSe$_2$ or WS$_2$, the 2D TMD dominated their characteristics, and this aligned with the absorption coefficient figures too. When polarized light in the x or y direction passed through the pristine and drug(H)/2D TMD complexes, the absorption behavior was investigated. The Pristine WSe$_2$ structure had the same absorption behavior in both directions, and absorption started from the onset of the bandgap. WSe$_2$ drug complexes also showed the same trend. For pristine WS$_2$ absorption started from the edge of the corresponding bandgap wavelength, whether y or x polarized light was incident on it. Drug(H)/ WSe$_2$ complexes showed a similar tendency.

The utilization of NIR photothermal therapy is not feasible for drug(H)/WS$_2$ structures because of the absence of a strong absorption peak within the NIR spectral range (700-1350 nm). Finally, it can be said that photothermal therapy can be achievable with both drug(H)/WS$_2$ and drug(H)/ WSe$_2$ structures, whether x or y polarized light was incident. However, NIR can be achievable only for drug(H)/WSe$_2$ structures. In this regard, WSe$_2$ drug complexes are in a better place because the use of NIR is highly appreciable in photothermal therapy because of better biological penetration of NIR light in the body \cite{zhang2021recent}.

\subsection{ Release of Drug Molecules }\label{subsec2}

\begin{figure*}[htbp]
    \centering
    \includegraphics[scale= 0.4]{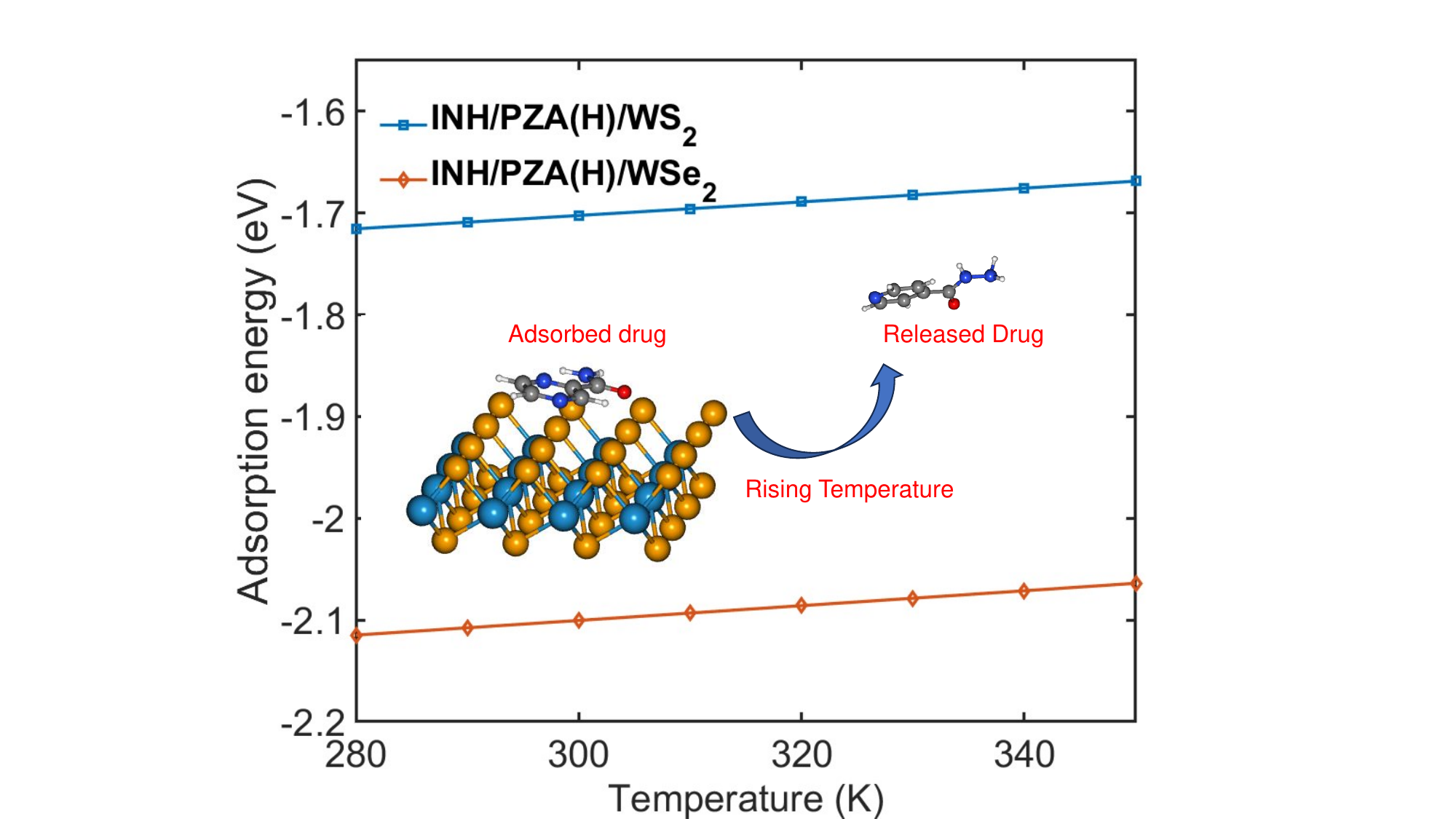}
     \caption{Temperature dependent adsorption energy of INH/PZA(H)/WS$_{2}$ and INH/PZA(H)/WSe$_{2}$ complexes }
    \label{adsorption}
\end{figure*}

After optical radiation is applied in photothermal therapy, releasing drug compounds from drug/2D TMD structures is critical in targeted drug delivery systems. Hence, adsorption should be a function of temperature in this case. The previous calculations of adsorption energy were done at T= 0 K with the idealized criteria and omitting the thermal vibrational component of the system. In temperature-controlled studies, such as drug delivery in photothermal treatment, it is imperative to consider the Helmholtz free energy (F) as the essential thermodynamic potential. Consequently, the thermal corrections in the adsorption energy were determined by considering the phonon vibration modes as quantum harmonic oscillators. Vibrational energy is temperature-dependent, and adding it to ideal adsorption energy will give a more accurate result. Vibrational energy was calculated with the “Finite Difference Method”.
\begin{equation}
F = E_{DFT} + A_{vibrational}
\end{equation}

Fig. \ref{adsorption} demonstrates the change of adsorption energy with temperature. Both drug/2D TMD complexes showed an increase in adsorption energy with the increase in temperature. INH/PZA(H)/WS$_2$ graph shows a 0.3034 eV increase in adsorption energy when T = 350 K compared with T = 0 K  where INH/PZA(H)/WSe$_2$ graph demonstrated a 0.2861 eV increase when T = 350K.The rise of adsorption energy will give a temperature-controlled release of anti-TB drug molecules. Since the formation energy (or adsorption energy of drug molecules) of 2D TMD complexes was rising, the probability of the anti-TB drug compound being isolated from weakly physisorbed interaction is higher. This method of temperature-dependent adsorption energy calculation was reported previously\,\cite{liang2020theoretical}. Moreover, this temperature-controlled releasing mechanism is experimentally proven for other 2D material-based targeted drug delivery systems. For example, a multifunctional MoS$_2$-based targeted drug delivery system for tumor cells was successfully synthesized, and the drug was released with photothermal therapy in the NIR region \cite{release}.

\section{Conclusion}\label{sec13}
We systematically investigated the interaction between two anti-TB drugs and two nanocarriers, WS$_2$ and WSe$_2$ monolayers, using DFT calculations. Our results suggested that all the drug/2D TMD complexes were stable because of their favorable adsorption energy. The most stable structure was found when INH and PZA drugs were simultaneously adsorbed on a TMD sheet. Further analysis of bands and DOS suggested that band structures were mainly dominated by pristine TMD materials, which were more comprehensively understood by projected band structures. Moreover, EDD and Mulliken charge analysis of the drug(H)/TMD complexes ensured no chemical bond was formed between drug atoms and 2D TMD materials. This suggests that both drugs can be easily released to the targeted site. Optical analysis showed that drug(H)/WS$_2$ complexes had good absorption within the visible range and drug(H)/WSe$_2$ had good absorption peaks within the NIR range. This shows the possibility of using photothermal therapy with our drug delivery system. Finally, we showed the temperature-dependent releasing mechanism of our TMD material complexes and ensured that the discussed anti-TB drugs could be released by creating heat through photothermal therapy. Our proposed targeted drug delivery system consisting of WS$_2$ or WSe$_2$ sheet showed great promise regarding their potential applications in photothermal therapy and temperature-dependent releasing behavior. The integration of combined photothermal therapy and chemotherapy treatments could revolutionize the field of TB therapy, paving the way for more effective and tailored approaches to combat this deadly disease.

\section*{Acknowledgments}

K.M. and T.Y. acknowledge the Nanoscale Simulation, Characterization and Fabrication Lab, Department of EEE, BUET, supervised by A.Z., for this work. All the authors thank the Department of EEE, BUET for providing the necessary support. T.Y acknowledges the funding from the Research and Innovation Centre for Science and Engineering (RISE), BUET.

\section*{Author Contributions}
\textbf{Khaled Mahmud}: Conceptualization, Formal analysis, Methodology, Visualization, Software, Investigation, Writing - original draft. \textbf{Taki Yashir }: Conceptualization, Methodology, Visualization, Software, Investigation, Writing – original draft. \textbf{Ahmed Zubair}: Supervision, Conceptualization, Methodology, Visualization, Project administration, Resources, Writing - original draft, Writing-review \& editing. 

\section*{Conflicts of interest}
There are no conflicts to declare.



\balance


\bibliography{TB_drug_delivery} 
\bibliographystyle{rsc} 

\end{document}